# Behavioral acceptance of automated vehicles: The roles of perceived safety concern and current travel behavior

Fatemeh Nazari [a,*], Mohamadhossein Noruzoliaee [a], Abolfazl (Kouros) Mohammadian [b]

[a] Department of Civil Engineering, University of Texas Rio Grande Valley, USA
[b] Department of Civil, Materials, and Environmental Engineering, University of Illinois at Chicago, USA

**Abstract:** With the prospect of next-generation automated mobility ecosystem, the realization of the contended traffic efficiency and safety benefits are contingent upon the demand landscape for automated vehicles (AVs). Focusing on the public *acceptance behavior* of AVs, this empirical study addresses two gaps in the plethora of travel behavior research on identifying the potential determinants thereof. First, a clear behavioral understanding is lacking as to the *perceived concern about AV safety* and the consequent effect on AV acceptance behavior. Second, how people appraise the benefits of enhanced automated mobility to meet their *current (pre-AV era) travel behavior* and needs, along with the resulting impacts on AV acceptance and perceived safety concern, remain equivocal. To fill these gaps, a recursive trivariate econometric model with ordinal-continuous outcomes is employed, which jointly estimates AV acceptance (ordinal), perceived AV safety concern (ordinal), and current annual vehicle-miles traveled (VMT) approximating the current travel behavior (continuous). Importantly, the co-estimation of the three endogenous outcomes allows to capture the true interdependencies among them, net of any correlated unobserved factors that can have common impacts on these outcomes. Besides the classical socio-economic characteristics, the outcome variables are further explained by the latent preferences for vehicle attributes (including vehicle cost, reliability, performance, and refueling) and for existing shared mobility systems. The model estimation results on a stated preference survey in the State of California provide insights into proactive policies that can popularize AVs through gearing towards the most affected population groups, particularly vehicle cost-conscious, safety-concerned, and lower-VMT (e.g., travel-restrictive) individuals.

---

[*] Corresponding author.
  *Email addresses:* fatemeh.nazari@utrgv.edu (F. Nazari), h.noruzoliaee@utrgv.edu (M. Noruzoliaee), kouros@uic.edu (A. Mohammadian)

# 1. Introduction

## 1.1. Motivation

The unparalleled technological advances in vehicle automation and artificial intelligence have made self-driving or automated vehicles (AVs) technically available for extensive road tests (e.g., Liu et al. (2019) and Zoellick et al. (2019)) and, even recently, for limited commercial shared mobility services (Waymo, 2020). By disengaging users from the driving task, AVs allow them to engage in desired non-driving activities, while also providing door-to-door mobility through self-parking capability, enhanced transit for persons with travel-restrictive conditions, improved accessibility to public transportation by serving as first/last mile mode, and traffic congestion and emissions mitigation through vehicle-to-vehicle and -infrastructure communications (Fagnant and Kockelman, 2015; Andreson et al., 2016; Harper et al., 2016; Meyer et al., 2017). These benefits to societies can be fully realized if AVs are widely accepted and in turn adopted by the public. However, people are observed to be mostly reluctant or neutral towards AVs at least in the near future, according to a recent review study (Harb et al., 2021) and empirical findings from the literature (Schoettle and Sivak, 2014; Menon et al., 2016; Haboucha et al., 2017; Puget Sound Regional Council, 2019). The public's negative tendency towards AVs likely leads to an associated slow market penetration (Chen et al., 2016; Noruzoliaee, 2018).

To identify factors negatively influencing the public's interest in AVs, the present study explores the individuals' *behavioral intention to accept AVs*, whereby the findings can elucidate the potential determinants of AV acceptance behavior in terms of individuals' characteristics and attitudinal profiles. By recognizing heterogeneous cohorts of individuals behaving differently regarding AV acceptance, the findings from this empirical study can further provide policy makers and stakeholders with incentivizing policy recommendations effectively geared towards appropriate population groups. In a broader perspective, the in-demand research question on whether and how the public accepts AVs is essential in the *travel behavior* domain, as also emphasized in a review study by Becker and Axhausen (2017). It not only is tightly related but also a fundamental input to various aspects of travel behavior, such as valuation of in-vehicle travel time (Steck et al., 2018; Rashidi et al., 2020; Szimba and Hartmann, 2020), possible changes made by AVs in individuals' future activity-travel behavior (e.g., mode choice (Yap et al., 2016; Malokin et al., 2019), activity pattern (Pudāne et al., 2019), and residential and work relocations (Krueger et al., 2019; Kim et al., 2020; Moore et al., 2020)), and urban planning and the built environment (Fraedrich et al., 2019). The comprehensive reviews of the automated travel behavior-related works can be found in Becker and Axhausen (2017), Gkartzonikas and Gkritza (2019), Soteropoulos et al. (2019), Keszey (2020), and Harb et al. (2021).

As a potential game changer that could transform the mobility safety, it is of paramount importance to scrutinize AV acceptance behavior vis-à-vis the safety aspect of AVs. On the one hand, the public could perceive safer mobility with AVs (Bansal et al., 2016) due to elimination of human drivers, whose driving error is the leading cause of nearly 94 percent of the U.S. road accidents (National Highway Traffic Safety Administration, 2019). One the other hand, the public could be concerned about AV safety due to potential lurking failures and malfunctions in unexpected weather/road conditions and cyber-attacks (Gomes, 2014; Kyriakidis et al., 2015; Andreson et al., 2016; Favarò et al., 2017), as also exemplified by crashes in AV road tests (Digital Trends, 2018). As it turns out in the empirical results of the earliest surveys and polls (Howard and Dai, 2014; Kyriakidis et al., 2015; Bansal et al., 2016) through the recent ones (Panagiotopoulos and Dimitrakopoulos, 2018; Liljamo et al., 2018; Puget Sound Regional Council, 2019), the public remains consistently concerned about safety issues of AVs. More importantly, the concern about AV safety is found to be a barrier to AV acceptance (Yap et al., 2016; Xu et al., 2018; Nazari et al., 2018; Zhang et al., 2019; Kassens-Noor et al., 2020; Dannemiller et al., 2021).

In light of this, the research efforts exploring how AV safety concern influences AV acceptance behavior need to be extended in order to also investigate what factors shape the public's *perception* of AV safety concern in terms of, for instance, personal characteristics and attitudes. This research need is also highlighted in a recent study (Wang et al., 2021), suggesting to further incorporate barriers such as AV safety issues into the behavioral studies. The first contribution of the present study is advancing this knowledge through a *joint modeling framework* to unravel the impact on AV acceptance of individuals' perceived concern about AV safety, among other influential factors, while at the same time "endogenously" connecting the perceived safety concern to the individuals' characteristics and attitudinal profiles. Notably, the joint model can disentangle the true effect of AV safety concern on AV acceptance from the effect of any unobserved factors that commonly influence both AV safety concern and AV acceptance behavior (i.e., endogeneity effects).[1] As discussed further in section 3, accommodating the endogeneity issue could help avoid inconsistent estimation results and in turn misleading policy recommendations.

Furthermore, a growing but limited number of studies on AV acceptance behavior looks into the associated role of the individuals' *current (pre-AV era)* travel behavior (Becker and Axhausen, 2017). This is motivated by the fact that the difference in the tendency of individual cohorts towards the AV technology may stem from their distinct travel needs and constraints, which could lead to dissimilar perceptions of the automated mobility gains. The current travel behavior factor is usually represented by driving frequency (Rödel et al., 2014) or by vehicle-miles traveled (VMT) (Kyriakidis et al., 2015; Bansal et al., 2016; Nazari et al., 2018), as categorized in a review of behavioral studies (Becker and Axhausen, 2017). While providing valuable insights, previous research considers the current travel behavior as an "exogenous" factor, which may lead to misleading estimation results. To tackle, the second contribution of this paper is advancing the understanding of the impact of current travel behavior (in terms of current annual VMT) on AV acceptance, while also explicitly accounting for the possible endogeneity effect through extending the joint modeling framework discussed in the previous paragraph. Specifically, the framework explores how AV acceptance behavior differs in diverse cohorts of individuals with different current travel behaviors and, at the same time, "endogenously" explains the individuals' current travel behavior based on their demographic characteristics and attitudes.

---

[1] Readers interested in exhaustive discussions on endogeneity are referred to Greene (2000) for linear regression models and to Ben-Akiva et al. (2002), Raveau et al. (2010), and Bhat and Dubey (2014) for choice models.

Another intriguing yet underexplored research question is whether and how a person's current travel behavior influences his/her perception of concern about AV safety. For example, higher-VMT individuals may be more safety-concerned (because of their potentially frequent/longer presence in vehicle, leading to a higher chance of experiencing an AV malfunction) or less concerned (since frequent/longer presence in vehicle may increase the chance of human driver fatigue and error). While a clear answer is lacking in the related work, the underline{third} contribution of this study is attempting to shed light on this question by assessing a hypothesis that an individual's current travel behavior affects his/her perception of AV safety concern.

**1.2. The present study in the context**

To fill the above-discussed research gaps, this study presents a recursive trivariate model with ordinal-continuous outcomes. The three outcome variables are defined as the level of intention to accept AVs (ordinal), the level of perceived concern about AV safety (ordinal), and the current travel behavior approximated by personal annual VMT (continuous). The "recursive" structural system of the three equations allows to capture the structural effects of the outcomes on each other, including the effects of: 1) AV safety concern on AV acceptance, 2) VMT on AV acceptance, and 3) VMT on AV safety concern. In addition, the "joint" estimation of the three outcomes enables to uncover any correlated unobserved factors commonly influencing the outcomes (i.e., endogeneity effects arising from simultaneity). Accommodating these structural endogeneity effects are crucial to finding the "true" structural relationships among the three endogenous outcomes, which are purified from the influence of common unobserved factors on the outcomes.

The three outcome variables are explained by the classical explanatory factors, such as socio-demographic characteristics, and unobservable (latent) subjective constructs reflecting individuals' preferences and attitudes. The incorporation of latent constructs into behavioral studies on AVs is highlighted as a research need in a recent study (Wang et al., 2021), and in general results in a more behaviorally realistic results ((Ben-Akiva et al., 2002), and refer to Lavieri et al. (2017) and Nazari et al. (2019) for empirical research examples). Through one group of the latent constructs, we attempt to explore the associated impacts of individuals' preferences for four vehicle attributes, including vehicle cost, reliability, performance, and refueling, given that the study in the context is a vehicle-related decision. One way to tackle this is through choice experiments given alternative attributes. Instead, we seek for a general understanding of the individuals' preferences through the latent constructs, partly because AVs have yet to enter the vehicle market, and thus the public's perception of their attributes is blurred. The other group of the latent constructs is built on the individuals' preferences for the existing on-demand mobility technologies (i.e., carsharing and ridesharing), which may partially explain mobility tech-savviness. The five latent constructs are built on the underlying self-reported indicators by estimating a structural equation model (SEM) with latent variables which precedes the full information maximum likelihood estimation of the recursive trivariate model. The empirical estimation of the model using a dataset collected in the State of the California results in valuable insights for policy decisions.

The remainder of the paper is organized as follows. The relevant background is reviewed in section 2. Section 3 presents formulation of the recursive trivariate model with latent constructs. The dataset for model estimation and the results are discussed in sections 4 and 5, respectively. The paper concludes in section 6 with a summary of the research findings, policy implications, and directions for future research.

## 2. Literature review

Over the past decade, there has been a growing interest in analyzing the transportation system landscape with the emergence of AV technology. On the demand side, several aspects of the users' responses to AVs are explored, which are thoroughly reviewed in a few studies. Starting with Becker and Axhausen (2017), the authors review studies asking a sample of individuals' general opinion about whether, when, how, and why they accept AVs, and then relating the individuals' responses (as outcome variables) to their socio-demographic characteristics, current travel behavior, trip attributes, as well as their concerns and attitudes (as explanatory variables). The authors furthermore comparatively analyze the findings of the reviewed studies on the effects of the explanatory variables on the outcome variables. By reviewing prior work on AVs using stated preferences and choice models, Gkartzonikas and Gkritza (2019) categorizes them based on the study objective, methodology, and sample population. The authors conclude that the majority of previous research aims at capturing individuals' behavioral characteristics and perception or willingness-to-pay for AV use. Moreover, the authors discuss the benefits, barriers/concerns, and opportunities of AV deployment.

More recently, Harb et al. (2021) review studies investigating AV implications on travel behavior and categorize them, based on the method of analysis, into controlled testbeds, driving simulators and virtual reality, agent-based and travel-demand models, surveys, and field experiments. Moreover, regarding the research questions, the authors categorize the studies into five groups. In the first group, the trust in and intention to accept/adopt/use AVs are investigated through economic approach (Yap et al., 2016; Haboucha et al., 2017; Wang and Zhao, 2019; Wang et al., 2021) and psychological approach (Hohenberger et al., 2016; Rahman et al., 2019). The research studies in the second and third groups analyze individuals' in-vehicle behavior, such as engagement in on-board activities (Pudāne et al., 2019), and valuation of travel time (Steck et al., 2018; Rashidi et al., 2020; Szimba and Hartmann, 2020), respectively.

The fourth group encompasses studies probing into the impacts of AVs on the individuals' travel behavior in the short- and long-term. For instance, Kim et al. (2019) explore mode use, which is a short-term travel behavior, across individuals with AV-enthusiast, AV-resistant, and anti-AV attitudes built as latent constructs on their perception of AV advantages and disadvantages by estimating a latent class cluster analysis on a dataset collected in the State of Georgia. In another example, Dannemiller et al. (2021) estimate a multivariate ordinal probit on a sample dataset collected in the Austin area, Texas, to investigate the AV effects on short-term activity-travel decisions, including trip generation and distance to destinations in local area for shopping/eating-out or leisure activities as well as making long-distance road trips and commute travel time. An example of analyzing the long-term AV impacts on travel behavior is the study of Fraedrich et al. (2019) on built environment through systematically reviewing the literature along with conducting a quantitative online survey and qualitative interviews with representatives from urban transport planning authorities in Germany. In the last group of studies, the effect of AVs on the individuals' VMT is modeled using agent-based network analysis (Fagnant and Kockelman, 2014, 2018), activity-based models (Auld et al., 2017; Kröger et al., 2019), and field experiments (Harb et al., 2018). Also, Soteropoulos et al. (2019) review and comparatively analyze findings of the prior works on the AV impacts on travel behavior — in terms of vehicle miles and hours traveled and mode share — as well as on land-use — in terms of parking space and location choice.

On examining how the AV acceptance behavior is influenced by concern about safety aspects of AVs and what factors shape perception of the safety concern, which is also one focus of this paper, one stream of studies with *psychological* perspective looks into personal concerns, perceptions, and attitudes by building latent psychometric constructs (see Keszey (2020) for a review of studies with psychological approach). The modeling frameworks of the psychological studies are built on theories such as the Technology Acceptance Model (Davis, 1989; Davis et al., 1989), Theory of Planned Behavior (Ajzen, 1991), and Unified Theory of Acceptance and Use of Technology (Venkatesh et al., 2003, 2012), by applying structural equation modeling technique on experimental or empirical datasets.

In experimental settings, subjects are first exposed to an AV ride in a road test or a driving simulator, and then inquired about their psychometric indicators about AV safety and their AV acceptance (e.g., Du et al. (2019), Zoellick et al. (2019), and Paddeu et al. (2020)). For instance, Xu et al. (2018) assess the impact of direct experience of 300 students with an AV ride on their intention to accept AVs by extending the Technology Acceptance Model, wherein they find that the safety concern factor directly and positively affects AV acceptance. In empirical studies, subjects respond to questionnaires that provide a description of AVs, followed by questions about psychometric indicators and AV acceptance (e.g., Hulse et al. (2018), Lee et al. (2019), and Zhang et al. (2019)). An example is the study by Zhang et al. (2019), who propose a new extension of the Technology Acceptance Model by including latent constructs explaining individuals' perception of trust, safety risk, and privacy risk. Implementing the model on a sample of 216 drivers, collected via a face-to-face interview, the authors report that a positive attitude towards AVs is mostly affected by initial trust which can be improved by reducing perceived safety risk associated with AVs and if users find AVs to be useful. Comparing the experimental and empirical approaches, Walker et al. (2019) find a strong relationship between the electrodermal activity, which can capture real-time changes in drivers' trust, and the self-reported trust in questionnaires.

Compared to the psychological studies, which capture the heterogeneity across the individuals by "exogenously" segmenting them post-estimation into groups with different socio-economic characteristics, another line of research with *economic* approach "endogenously" relates the intention to accept AVs with the classical attributes of individual decision makers, such as socio-demographics and current travel behavior, as well as latent psychometric constructs explaining individuals' concerns, attitudes, and perceptions (reviews of these research works can be found in Becker and Axhausen (2017), Gkartzonikas and Gkritza (2019), and Harb et al. (2021)). The majority of the economic studies, as reviewed below, apply the well-established discrete choice theory (McFadden, 1981) on datasets collected through stated preferences surveys to analyze the relationship between AV acceptance and concerns about AV safety issues.

Among those investigating the AV safety concern, Yap et al. (2016) present a sequential model, where confirmatory factor analysis is used to build a latent construct explaining trust in the safety of a trip with an AV, which is subsequently plugged into a mixed logit choice model to evaluate preference for AV as a last-mile mode. Estimating the model on a stated preferences dataset collected in the Netherlands, the authors find that trust, which negatively affects AV attractiveness, is the second-largest contributor to the utility of traveling with AVs. Also, Nazari et al. (2018) utilize a two-step model, wherein a structural equation model is first estimated to relate the latent concern about AV safety to the associated indicators and personal characteristics. The latent construct is subsequently fed into a multivariate model in the second step to elicit the public interest in both private and shared AVs for both daily and commute trips using a stated preferences dataset collected in the U.S. The most impeding factor for interest in AVs was found to be the safety concern.

In a more recent study, Wang and Zhao (2019) developed a sequential, three-step model to analyze the relationship between risk preference and adoption of AVs. With risk preference measured in the first step based on both the prospect theory and factor analysis, the second step ties the measured preference with socio-economic characteristics using linear regression, which was forwarded to the third step exploring the preferences for AVs in a mixed logit model. Also, estimating a logistic regression on a database collected in the State of Michigan, Kassens-Noor et al. (2020) report that almost half of the current public transit users were reluctant to use autonomous buses, mainly due to their main concern about the ride safety. Recently, Dannemiller et al. (2021) applies a confirmatory factor analysis to construct latent variables, including AV safety concern, which are then given to five multivariate ordinal-outcome models to analyze the associated impacts on five dimensions of short-term activity-travel choices. Estimated on a dataset collected in Austin the authors find safety concern as a dominant reason for low tendency towards AVs, mostly found in persons who are likely characterized as older, females, low-income, unemployed, and those with children.

Another focus of the present study is on how AV acceptance behavior is influenced by current travel behavior usually represented by annual VMT, driving frequency, car availability, or multimodality, which are thoroughly reviewed in Becker and Axhausen (2017). A prior research example is the study of Kyriakidis et al. (2015), who conduct an online worldwide survey to analyze the public opinion on acceptance of, concerns about, and willingness-to-buy AVs. By calculating Spearman correlation, the authors find that driving more, in terms of mileage and driving frequency, leads to higher willingness-to-pay for AVs, likely due to more time spent in vehicles. Also, Bansal et al. (2016) estimate ordinal probit models using a sample dataset collected in Austin to investigate willingness-to-pay for levels of automation and connectivity added to the individuals' current as well as next vehicles. They conclud that those who driver more, represented by higher annual VMT, are more willing to pay for AVs with higher automation levels. Moreover, Krueger et al. (2016) collect a sample dataset in Australia to investigate individuals' mode choice among on-demand AV with and without ride-sharing and public transit. By analyzing results of a mixed logit model, the authors report that individuals' current travel behavior as captured by multimodality increases the probability of adopting AVs.

# 3. Methodology

Figure 1 outlines the two-step modeling framework estimated sequentially.[2] The first step (section 3.1) estimates an SEM with latent variables to identify latent preferences for vehicle attributes and shared mobility. In the second step (section 3.2), we present a recursive trivariate model with ordinal-continuous outcomes, which jointly estimates three dependent variables shown by the yellow boxes in Figure 1 as well as the structural effects of these outcomes on each other (represented by the solid red and blue arrows in Figure 1).

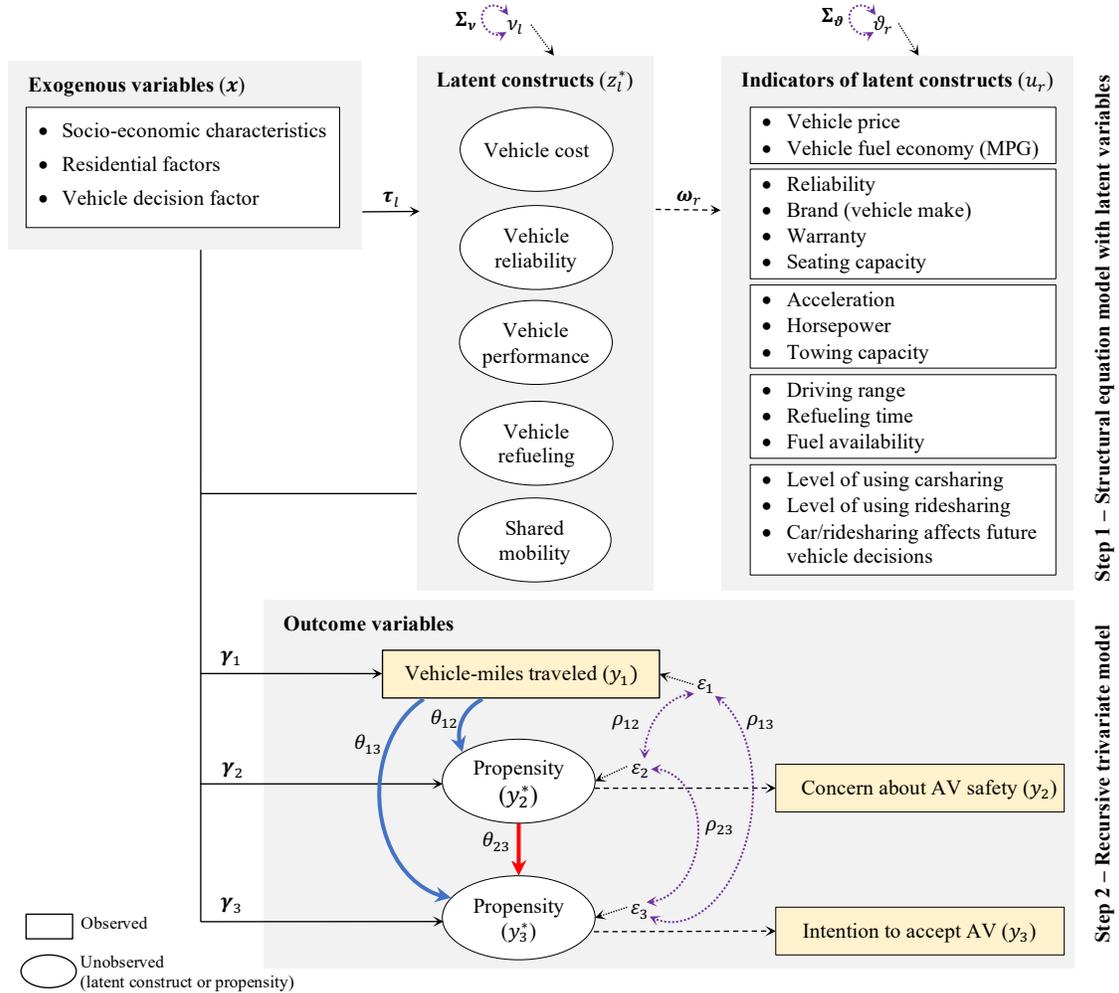

**Figure 1. The modeling framework**

---

[2] While the simultaneous estimation of the two steps may improve the results at a higher computational cost, the consequent improvement compared to sequential estimation has shown to be small in previous empirical studies (e.g., Raveau et al. (2010)). Moreover, the presence of two ordinal outcomes in the second step of this model leads to even higher computational cost, due to the appearance of several integrations in the likelihood function, than that of a model with discrete outcome (e.g., an integrated multinomial logit with latent variables). Besides, the sequential estimation can theoretically yield acceptable results when the sample dataset is large enough (Ben-Akiva et al., 2002).

## 3.1. Structural equation model with latent variables

For each individual $i \in \{1, 2, \ldots, N\}$, the SEM with latent variables comprises two sets of measurement and structural equations. For notational convenience, the index for individuals is suppressed in the rest of this section. The measurement equations (1) relate a vector of latent preferential constructs $\boldsymbol{z}^* = (\ldots, z_l^*, \ldots)'$, $l \in \{1, \ldots, L\}$, to the underlying observed indicators $u_r, r \in \{1, \ldots, R\}$, through the associated vector of estimable loadings $\boldsymbol{\omega}_r$. This is represented in Figure 1 by the dashed arrow from the box of latent constructs towards the box of observed indicators. Simultaneously, the structural equations (2) explain each latent construct $z_l^*$ using a vector of exogenous variables $\boldsymbol{x}$ and the associated vector of estimable coefficients $\boldsymbol{\tau}_l$. This is also depicted in Figure 1 by the solid arrow from the box of exogenous variables towards the box of latent constructs. The effects of unknown or omitted factors are captured by the random error terms $\vartheta_r$ and $v_l$ in Eqs. (1)-(2) (shown by the two-sided dotted purple arrows in the upper part of Figure 1), which are assumed to be standard multivariate normally distributed, i.e., $\vartheta_r \sim [0, \boldsymbol{\Sigma}_\vartheta]$ and $v_l \sim [0, \boldsymbol{\Sigma}_v]$, wherein $\boldsymbol{\Sigma}_\vartheta$ and $\boldsymbol{\Sigma}_v$ denote the covariance matrices. The system of equations (1)-(2) is solved using the maximum likelihood method to estimate the model parameters.

$$u_r = \boldsymbol{\omega}_r' \boldsymbol{z}^* + \vartheta_r \qquad r \in \{1, \ldots, R\} \tag{1}$$

$$z_l^* = \boldsymbol{\tau}_l' \boldsymbol{x} + v_l \qquad l \in \{1, \ldots, L\} \tag{2}$$

## 3.2. Recursive trivariate model with ordinal-continuous outcomes

For each individual, the recursive trivariate model jointly estimates one linear regression equation with continuous outcome (i.e., logarithm of current annual VMT) and two equations with ordinal outcomes (i.e., level of concern about AV safety and level of intention to accept AVs). Eq. (3) models the continuous VMT outcome, $y_1$, as a linear function of the vectors of exogenous variables $\boldsymbol{x}$ (including socio-economic, residential, and vehicle decision factors) and expected values of the latent variables $\hat{\boldsymbol{z}}^* = (\ldots, \hat{z}_l^*, \ldots)'$ estimated in section 3.1. For notational simplicity, we stack the explanatory variables ($\boldsymbol{x}$ and $\hat{\boldsymbol{z}}^*$) and their estimable coefficients ($\boldsymbol{\alpha}_1$ and $\boldsymbol{\beta}_1$) into vectors $\boldsymbol{w}$ and $\boldsymbol{\gamma}_1$, respectively. The unobserved error term of the linear equation, $\varepsilon_1$, is assumed to be normally distributed. It is assumed that the explanatory variables meet the condition of strict exogeneity, i.e., $E(\varepsilon_1|\boldsymbol{w}) = 0$.

Eqs. (4)-(5) relate an individual's stated ordinal levels of agreement with being concerned about AV safety $y_2 = j_2 \in \{1, \ldots, J_2\}$ and with AV acceptance $y_3 = j_3 \in \{1, \ldots, J_3\}$ to the associated continuous latent propensities $y_2^*$ and $y_3^*$ through the estimable threshold parameters $\mu_{j_2}$ and $\mu_{j_3}$. In this paper, the ordinal levels range from 1 (strongly disagree) to 5 (strongly agree). The relationship is established through the exogenous variables $\boldsymbol{x}$, where their respective coefficient vectors are $\boldsymbol{\alpha}_2$ and $\boldsymbol{\alpha}_3$, and the expected values of the latent preferential constructs $\hat{\boldsymbol{z}}^*$, where their respective coefficient vectors are $\boldsymbol{\beta}_2$ and $\boldsymbol{\beta}_3$. The unobserved error components of the two equations, $\varepsilon_2$ and $\varepsilon_3$, are assumed to be standard normally distributed. The explanatory variables meet the conditions of strict exogeneity, i.e., $E(\varepsilon_2|w) = 0$ and $E(\varepsilon_3|w) = 0$, where $w$ is the stacked vector of $x$ and $\hat{z}^*$.

The joint estimation of Eqs. (3)-(5) relaxes the assumption of zero correlations between the unobserved error terms $\varepsilon_1$, $\varepsilon_2$, and $\varepsilon_3$. The underlying covariance matrix is written in Eq. (6), where $\rho_{12}$, $\rho_{13}$, and $\rho_{23}$ denote the correlations of the error terms, as shown by the two-sided dotted purple arrows in the lower part of Figure 1, and $\sigma_1$ represents the standard deviation of the error term $\varepsilon_1$. Importantly, the joint estimation further allows to capture the presence of endogeneity stemming from the interdependencies among the three outcome variables, which can lead to the presence of correlation between the endogenous outcome variables and the error terms. Specifically, we consider the endogeneity problem arising from the structural effects of VMT on both AV safety concern and AV acceptance, i.e., $E[\varepsilon_2|y_1] \neq 0$ and $E[\varepsilon_3|y_1] \neq 0$, as well as from the structural effect of AV safety concern on AV acceptance, i.e., $E[\varepsilon_3|y_2^*] \neq 0$.[3] The joint model can absorb any propensity for the outcomes caused by omitted or unobserved factors that have common effects on the outcomes. Consequently, the pure structural effects of the outcomes on each other, which are purified from the influence of common unobserved factors on the outcomes, can be obtained through the recursive nature of Eqs. (4)-(5). That is, the structural effects of VMT on the continuous latent propensities for AV safety concern and AV acceptance are captured by the recursive appearance of $y_1$ in Eqs. (4)-(5), with the corresponding coefficients denoted by $\theta_{12}$ and $\theta_{13}$, and represented by the solid blue arrows in Figure 1. Also, the structural effect of AV safety concern on AV acceptance is taken into account by the presence of $y_2^*$ and its coefficient $\theta_{23}$ in Eq. (5), as shown by the solid red arrow in Figure 1.

$$y_1 = \underbrace{\alpha_1'x + \beta_1'\hat{z}^*}_{\gamma_1'w} + \varepsilon_1 \tag{3}$$

$$y_2^* = \underbrace{\alpha_2'x + \beta_2'\hat{z}^*}_{\gamma_2'w} + \theta_{12}y_1 + \varepsilon_2 \qquad y_2 = j_2 \text{ if } \mu_{j_2-1} < y_2^* < \mu_{j_2}, \quad j_2 \in \{1,\dots,J_2\} \tag{4}$$

$$y_3^* = \underbrace{\alpha_3'x + \beta_3'\hat{z}^*}_{\gamma_3'w} + \theta_{13}y_1 + \theta_{23}y_2^* + \varepsilon_3 \qquad y_3 = j_3 \text{ if } \mu_{j_3-1} < y_3^* < \mu_{j_3}, \quad j_3 \in \{1,\dots,J_3\} \tag{5}$$

$$(\varepsilon_1, \varepsilon_2, \varepsilon_3) \sim N\left[\begin{pmatrix}0\\0\\0\end{pmatrix}, \begin{pmatrix}\sigma_1^2 & \rho_{12}\sigma_1 & \rho_{13}\sigma_1\\ \rho_{12}\sigma_1 & 1 & \rho_{23}\\ \rho_{13}\sigma_1 & \rho_{23} & 1\end{pmatrix}\right] \tag{6}$$

To estimate the recursive trivariate model, the likelihood function is derived based on the joint probability of the three outcomes, i.e., $P(y_1, y_2, y_3) = P(y_2, y_3|y_1) P(y_1)$. The marginal probability is immediately calculated as $P(y_1) = \frac{1}{\sigma_1}\phi\left(\frac{y_1-\gamma_1'w}{\sigma_1}\right)$, where $\phi$ is the standard normal density function. To derive the conditional probability $P(y_2, y_3|y_1)$, we start with writing the conditional distribution of the trivariate normal (Greene, 2000) as in Eq. (7).

---

[3] Strictly speaking, as the endogeneity problem can in general arise from any situation where an explanatory variable is correlated with an equation error term, it is emphasized herein that this empirical study accounts for the presence of endogeneity rooted only in simultaneity (i.e., co-estimation of multiple outcome variables, each of which can be affected by the other outcomes besides exogenous explanatory variables). Other sources of endogeneity, including omitted/unobserved confounding factors (wherein an omitted variable is correlated with both the equation error terms and exogenous explanatory variables) and measurement error (wherein exogenous explanatory variables are measured with uncertainty), are beyond the scope of this paper.

$$(\varepsilon_2, \varepsilon_3|\varepsilon_1) \sim N\left[\begin{pmatrix}\frac{\rho_{12}}{\sigma_1}(y_1-\gamma_1'w)\\ \frac{\rho_{13}}{\sigma_1}(y_1-\gamma_1'w)\end{pmatrix}, \begin{pmatrix}1-\rho_{12}^2 & \rho_{23}-\rho_{12}\rho_{13}\\ \rho_{23}-\rho_{12}\rho_{13} & 1-\rho_{13}^2\end{pmatrix}\right] \quad (7)$$

Rewriting Eqs. (4)-(5) as in Eqs. (8)-(9) gives the structural form as a recursive bivariate ordered probit model conditional on $y_1$. The structural form error components are denoted by $\eta_2$ and $\eta_3$, which have bivariate normal distribution with zero means and covariance matrix shown in Eq. (7).

$$y_2^*|y_1 = h_2 + \eta_2 \qquad y_2 = j_2 \text{ if } \mu_{j_2-1} < y_2^* < \mu_{j_2}, \ j_2 \in \{1, ..., J_2\} \quad (8)$$

$$y_3^*|y_1 = h_3 + \theta_{23}(y_2^*|y_1) + \eta_3 \qquad y_3 = j_3 \text{ if } \mu_{j_3-1} < y_3^* < \mu_{j_3}, \ j_3 \in \{1, ..., J_3\} \quad (9)$$

where, for notational convenience, the auxiliary variables $h_2$ and $h_3$ are defined as $h_2 = \gamma_2'w + \theta_{12}y_1 + \frac{\rho_{12}}{\sigma_1}(y_1-\gamma_1'w)$ and $h_3 = \gamma_3'w + \theta_{13}y_1 + \frac{\rho_{13}}{\sigma_1}(y_1-\gamma_1'w)$.

The solution of the conditional recursive bivariate ordered probit model in Eqs. (8)-(9) should determine $y_2^*$ in terms of $(w, y_1, \eta_2)$ and $y_3^*$ in terms of $(w, y_1, \eta_3)$, which represents the reduced form of the model and is shown in Eqs. (8) and (10).

$$y_3^*|y_1 = h_3 + \theta_{23}h_2 + \tilde{\eta}_3 \qquad y_3 = j_3 \text{ if } \mu_{j_3-1} < y_3^* < \mu_{j_3}, \ j_3 \in \{1, ..., J_3\} \quad (10)$$

where, $\eta_2$ and $\tilde{\eta}_3$ denote the reduced form error terms with bivariate normal distribution, as shown in Eq. (11). The covariance matrix in Eq. (11) is obtained by noting that $\tilde{\eta}_3 = \theta_{23}\eta_2 + \eta_3$, which implies $\text{var}(\tilde{\eta}_3) = \theta_{23}^2\text{var}(\eta_2) + 2\theta_{23}\text{cov}(\eta_2,\eta_3) + \text{var}(\eta_3)$ and $\text{cov}(\eta_2,\tilde{\eta}_3) = \theta_{23}\text{var}(\eta_2) + \text{cov}(\eta_2,\eta_3)$.

$$(\eta_1, \tilde{\eta}_2) \sim N\left[\begin{pmatrix}0\\0\end{pmatrix}, \begin{pmatrix}1-\rho_{12}^2 & \theta_{23}(1-\rho_{12}^2)+\rho_{23}-\rho_{12}\rho_{13}\\ \theta_{23}(1-\rho_{12}^2)+\rho_{23}-\rho_{12}\rho_{13} & \theta_{23}^2(1-\rho_{12}^2)+2\theta_{23}(\rho_{23}-\rho_{12}\rho_{13})+1-\rho_{13}^2\end{pmatrix}\right] \quad (11)$$

With the reduced form derived above, the joint probability of $y_2$ and $y_3$ conditional on $y_1$ can thus be written as in Eq. (12).

$$\begin{aligned}P(y_2=j_2, y_3=j_3|y_1) &= P(\mu_{j_2-1} < y_2^* \leq \mu_{j_2}, \mu_{j_3-1} < y_3^* \leq \mu_{j_3}|y_1)\\ &= P(y_2^* \leq \mu_{j_2}, y_3^* \leq \mu_{j_3} \ |y_1)\\ &\quad - P(y_2^* \leq \mu_{j_2-1}, y_3^* \leq \mu_{j_3} \ |y_1)\\ &\quad - P(y_2^* \leq \mu_{j_2}, y_3^* \leq \mu_{j_3-1}|y_1)\end{aligned} \quad (12)$$

$$+ P(y_2^* \leq \mu_{j_2-1}, y_3^* \leq \mu_{j_3-1}|y_1)$$

which gives:

$$
\begin{aligned}
P(y_2 = j_2, y_3 &= j_3|y_1) \\
&= \Phi_2\left((\mu_{j_2} - h_2)\lambda_2, (\mu_{j_3} - h_3 - \theta_{23}h_2)\lambda_3, \tilde{\rho}\right) \\
&- \Phi_2\left((\mu_{j_2-1} - h_2)\lambda_2, (\mu_{j_3} - h_3 - \theta_{23}h_2)\lambda_3, \tilde{\rho}\right) \\
&- \Phi_2\left((\mu_{j_2} - h_2)\lambda_2, (\mu_{j_3-1} - h_3 - \theta_{23}h_2)\lambda_3, \tilde{\rho}\right) \\
&+ \Phi_2\left((\mu_{j_2-1} - h_2)\lambda_2, (\mu_{j_3-1} - h_3 - \theta_{23}h_2)\lambda_3, \tilde{\rho}\right)
\end{aligned}
\quad (13)
$$

where $\Phi_2$ is the cumulative distribution function of the bivariate standard normal distribution. $\lambda_2, \lambda_3$, and $\tilde{\rho}$ are defined in Eqs. (14)-(16) and are used to normalize the reduced form error components in Eq. (11) to bivariate standard normal distribution.

$$\lambda_2 = \frac{1}{\sqrt{1-\rho_{12}^2}} \quad (14)$$

$$\lambda_3 = \frac{1}{\sqrt{\theta_{23}^2(1-\rho_{12}^2)+2\theta_{23}(\rho_{23}-\rho_{12}\rho_{13})+1-\rho_{13}^2}} \quad (15)$$

$$\tilde{\rho} = \lambda_2\lambda_3[\theta_{23}(1-\rho_{12}^2) + \rho_{23} - \rho_{12}\rho_{13}] \quad (16)$$

Putting together the derived conditional and marginal probabilities, the log-likelihood function for each observation (in this study, each observation refers to an individual) is formalized in Eq. (17). The model parameters can be estimated by maximizing the aggregated log-likelihood functions over a sample of $N$ individuals.

$$\ln L = \left(\sum_{j_2=1}^{J_2}\sum_{j_3=1}^{J_3} I(y_2 = j_2, y_3 = j_3) \ln P(y_2 = j_2, y_3 = j_3|y_1)\right) + \ln\left[\frac{1}{\sigma_1}\phi\left(\frac{y_1 - \boldsymbol{\gamma}_1'\boldsymbol{w}}{\sigma_1}\right)\right] \quad (17)$$

Once the model is estimated, the signs of the estimated coefficients in the equations of the two ordinal outcomes $y_q$, $q \in \{2,3\}$, can be used to interpret merely the highest ($y_q = J_q$) and the lowest ($y_q = 1$) ordinal levels. A positive (or negative) coefficient means that an increase in the corresponding explanatory variable increases (or decreases) the probability of $y_q = J_q$ and decreases (or increases) the probability of $y_q = 1$. The associated impacts on the intermediate ordinal levels ($y_q = 2, \ldots, y_q = J_q - 1$) can be determined through the marginal effect of each ordered level. To calculate the marginal effects, we assume a zero correlation between the error components $\varepsilon_q$ (Greene, 2000; Greene and Hensher, 2010; Washington et al., 2003). For simplicity, stack the coefficients ($\boldsymbol{\gamma}_2, \boldsymbol{\gamma}_3, \theta_{12}, \theta_{13}$, and $\theta_{23}$) and their explanatory variables ($\boldsymbol{w}$ and $y_1$) into vectors $\boldsymbol{\kappa}$ and $\boldsymbol{m}$, respectively. Eq. (18) calculates the marginal effects of the continuous explanatory variables $\boldsymbol{m}^c$ for ordinal level $\boldsymbol{j}_q$, wherein $\varphi$ denotes the probability density function of the normal distribution. For a dummy variable $\boldsymbol{m}^d$, Eq. (19) computes the marginal effect for ordinal level $\boldsymbol{j}_q$, which is equal to the effect of a change in $\boldsymbol{m}^d$ from 0 to 1 while all other exogenous variables are held at their arithmetic means (Greene, 2000; Greene and Hensher, 2010; Washington et al., 2003). The calculated values are then averaged over the sample of $N$ individuals.

$$ME_{j_q}(\boldsymbol{m}^c) = \frac{\partial P(y_q = j_q|\boldsymbol{m}^c)}{\partial \boldsymbol{m}^c} = \left[\varphi\left(\mu_{j_q-1} - \boldsymbol{\kappa}^c\boldsymbol{m}^c\right) - \varphi\left(\mu_{j_q} - \boldsymbol{\kappa}^c\boldsymbol{m}^c\right)\right]\boldsymbol{\kappa} \quad \forall q \in \{2,3\} \tag{18}$$

$$\begin{aligned}ME_{j_q}(\boldsymbol{m}^d) = &\left[\Phi\left(\mu_{j_q} - \boldsymbol{\kappa}^c\boldsymbol{m}^c + \boldsymbol{\kappa}^d\right) - \Phi\left(\mu_{j_q-1} - \boldsymbol{\kappa}^c\boldsymbol{m}^c + \boldsymbol{\kappa}^d\right)\right] \\ &- \left[\Phi\left(\mu_{j_q} - \boldsymbol{\kappa}^c\boldsymbol{m}^c\right) - \Phi\left(\mu_{j_q-1} - \boldsymbol{\kappa}^c\boldsymbol{m}^c\right)\right] \quad \forall q \in \{2,3\}\end{aligned} \tag{19}$$

## 4. Data

To estimate the presented recursive trivariate model with latent variables, we use a sample dataset of 3,574 individuals provided by National Renewable Energy Laboratory (2019). This section provides descriptive statistics of the sample dataset, encompassing the outcome variables (section 4.1), observed indicators of latent preferential variables (section 4.2), and exogenous variables (section 4.3).

### 4.1. Outcome variables

The respondents to the survey were inquired about their *stated* intention to accept an AV and *stated* concern about safety of AVs which form two ordinal outcomes of the recursive trivariate model (recalling from Figure 1). Specifically, the ordinal outcomes measure the individuals' agreement level with the following questions on a five-point Likert scale ranging from 1 (representing "strongly disagree") to 5 (representing "strongly agree").

(a) AV acceptance: "I would consider purchasing a vehicle that is fully self-driving (i.e., the vehicle drives itself)"

(b) AV safety concern: "I am concerned about the safety of self-driving vehicles"

In the first question, we assume that considering AV purchase can represent the intention to AV acceptance. The individuals' responses to the two questions are statistically distributed as Figure 2. Overall, more than one third of the individuals are inclined to accept AVs, whereas more than three quarters of them are concerned about safety of AV technology.

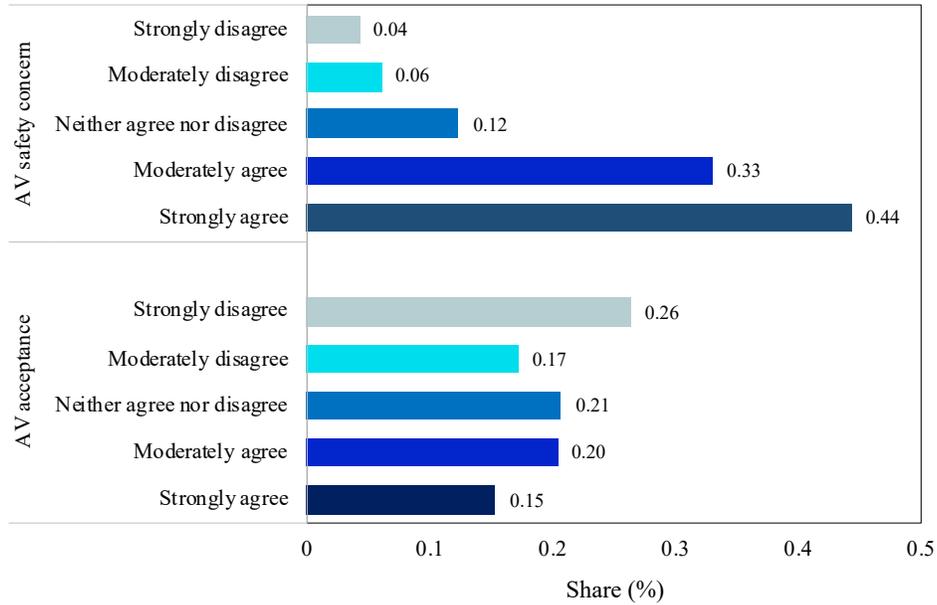

**Figure 2. Statistical distributions of the two ordinal outcome variables: stated intention to accept AV and stated concern about AV safety (sample size= 3,574)**

Furthermore, the survey respondents were inquired about their *revealed* VMT over the past year which is the continuous outcome of the recursive trivariate model, i.e., logarithm of annual VMT, (recalling from Figure 1). Figure 3 displays the (cumulative) distribution of the annual VMT over the sampled individuals in a Pareto diagram. Approximately, one third of the individuals drive 5,000 miles or less per year, while the majority of them drive 15,000 miles or less.

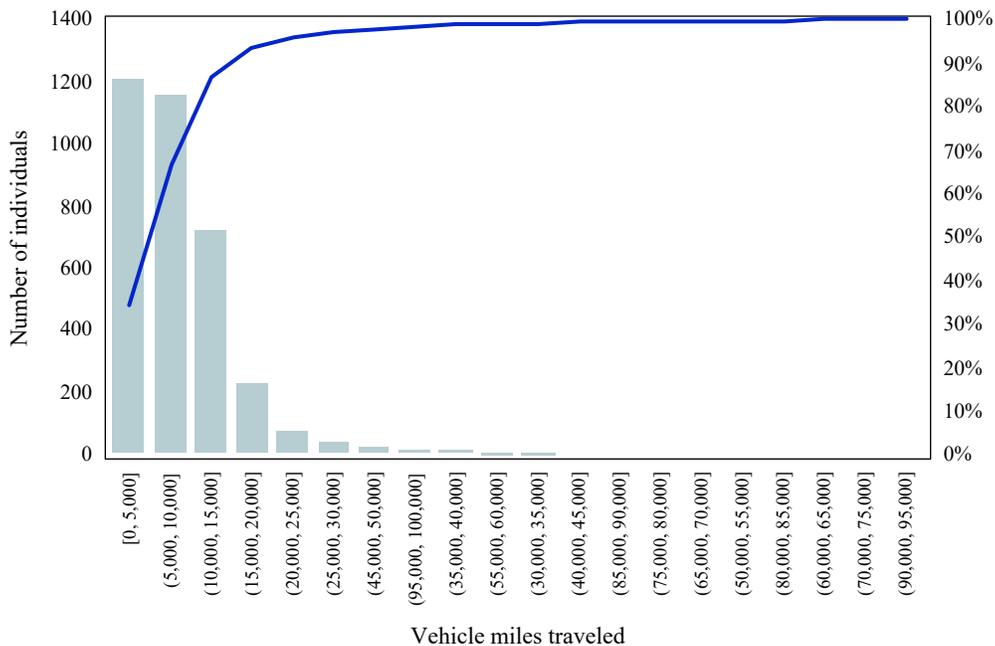

**Figure 3. Statistical distribution of the continuous outcome variable: revealed annual VMT (sample size = 3,574)**

## 4.2. Observed indicators of latent variables

As delineated in Figure 1, the SEM with latent variables yields five latent preferential constructs built on the underlying observed indicators, which are discussed in this section. The self-reported indicators measure the individuals' latent preferences for vehicle attributes — namely, vehicle cost, vehicle reliability, vehicle performance, and vehicle refueling — and shared mobility technologies. The individuals' responses to each indicator are averaged over the sample, as visualized in Figure 4.

Figure 4(a) itemizes twelve indicators associated with the latent preferences for four vehicle attributes: 1) vehicle price and fuel economy (in miles per gallon or MPG) are tied to the preference for vehicle cost, 2) reliability, brand, warranty, and seating capacity (which may represent the vehicle body size) correspond to the preference for vehicle reliability, 3) acceleration, horsepower, and towing capacity measure the preference for vehicle performance, and 4) driving range, refueling time, and fuel availability are associated with the preference for vehicle refueling. On average, the highest priorities are given to vehicle price, fuel economy, and reliability, whereas towing capacity, horsepower, and refueling time are expressed as the least preferred indicators.

According to Figure 4(b), the latent preference for shared mobility technologies is defined by three self-reported indicators. Two indicators measure the levels of participation in on-demand carsharing (e.g., Zipcar and Car2go) and ridesharing (e.g., Uber and Lyft) programs. On average, more than 60% of the individuals are somehow willing to participate in ridesharing, but more than half of them are not interested in carsharing. The third indicator relates to the expected impact of car/ridesharing programs on the individuals' future vehicle ownership decisions, with which only 20% of the sample agree.

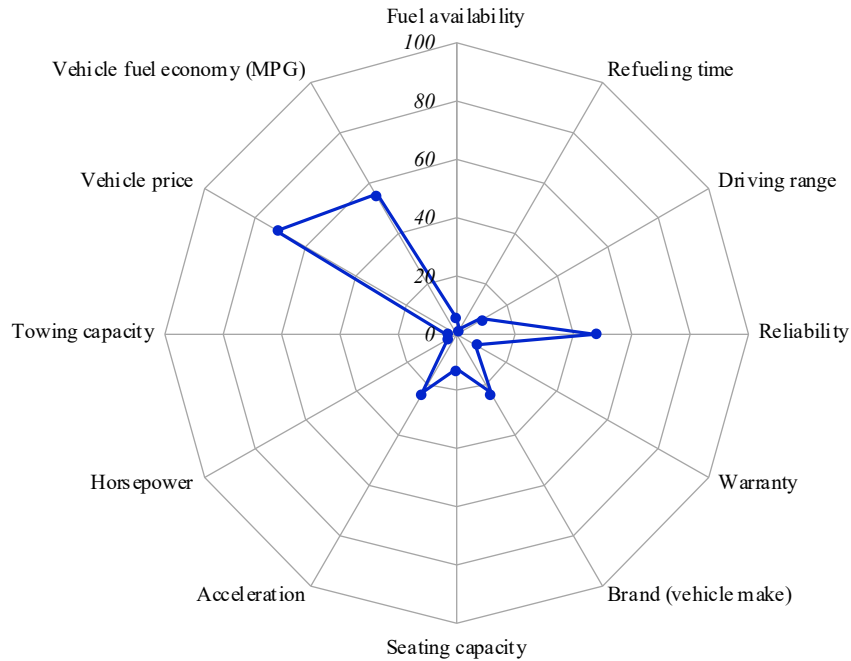

(a) Indicators of preferences for vehicle attributes

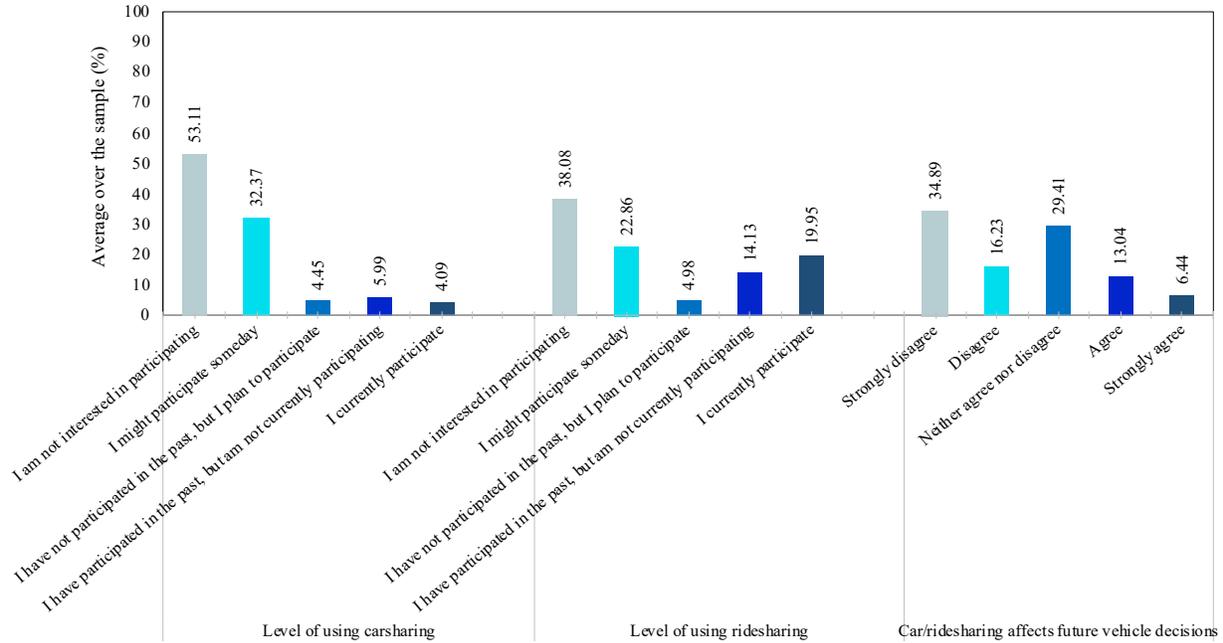

(b) Indicators of preferences for shared mobility

**Figure 4. Statistical distribution of the observed indicators of five latent preferential variables (sample size = 3,574)**

### 4.3. Exogenous variables

The exogenous variables describe the socio-economic, residential, and vehicle decision factors (Table 1). Among the socio-economic characteristics, the individuals are characterized by gender, race, education level, and employment type. The sample is almost equally distributed over males (49.61%) and females (50.39%). The sample distribution of race nearly follows the estimate by the U.S. Census Bureau (2018) for the State of California. Two thirds of the respondents are White, followed by 13.82% Asians, 3.22% African Americans, 1.15% American Indians or Alaska Natives, and 15.58% other race categories. Among four education levels, nearly two thirds of the individuals are either college graduates (4-year degree) or highly educated (post-graduate degree or higher). Regarding the employment type, the share of full-time employees, self-employed, and part-time employees are respectively 48.27%, 6.38%, and 10.97%. Furthermore, the household-level socio-economic attributes identify each individual by his/her household's structure and annual income. The former is defined by the number of children (age < 12), teenage children (12 ≤ age < 16), and adults (age ≥ 16). On average, the individuals live in households with two adult members. Three household income categories are low (< $75k), medium ($75k ≤ < $200k), and high (≥ $200k) levels with respective shares of 42.33%, 47.99%, and 9.68%.

Two residential factors describe the parking cost at residence (Table 1). The majority of the individuals park their vehicle(s) at their residence for free (for instance, at private garage or driveway). Averaged over the sample, each individual pays $0.502 per day (with a standard deviation equal to $7.336) for vehicle parking at residence. The last group of the exogenous variables describes how the individuals interact with their household members on the decision about future vehicle(s). As indicated in Table 1, more than two thirds are either sole decision-maker or primary decision-maker in their household, and the rest equally share their decision with other household member(s).

Table 1. Sample data for exogenous variables (sample size = 3,574)

| Exogenous variables | Category | No. of observations | Share (%) |
|---|---|---|---|
| *Socio-economic characteristics* | | | |
| Gender | Female | 1,801 | 50.39 |
|  | Male | 1,773 | 49.61 |
| Race | White | 2367 | 66.23 |
|  | Asian | 494 | 13.82 |
|  | African American | 115 | 3.22 |
|  | American Indian or Alaska Native | 41 | 1.15 |
|  | Other | 557 | 15.58 |
| Education | Level 1: High school graduate/GED or less | 245 | 6.86 |
|  | Level 2: Technical school/professional business school, some college, or community college graduate (associate 2-year degree) | 1,030 | 28.82 |
|  | Level 3: College graduate (4-year degree) | 1,160 | 32.46 |
|  | Level 4: Post-graduate degree or higher | 1,139 | 31.87 |
| Employment type | Full-time employed | 1,725 | 48.27 |
|  | Self-employed | 228 | 6.38 |
|  | Part-time employed | 392 | 10.97 |
|  | Not employed | 1,229 | 34.39 |
| Household structure | | | |
| No. of children (age < 12) | Mean = 0.318, Standard deviation = 0.715 | — | — |
| No. of teenage children ($12 \leq$ age < 16) | Mean = 0.105, Standard deviation = 0.362 | — | — |
| No. of adults (age $\geq$ 16) | Mean = 2.038, Standard deviation = 0.855 | — | — |
| Household income | Low level (< $75k per year) | 1,513 | 42.33 |
|  | Medium level ($75k $\leq$ < $200k per year) | 1,715 | 47.99 |
|  | High level ($\geq$ $200k per year) | 346 | 9.68 |
| *Residential factors* | | | |
| Free parking at residence | Yes | 3,336 | 93.34 |
|  | No | 238 | 6.66 |
| Parking cost at residence ($/day) | Mean = 0.502, SD = 7.336 | — | — |
| *Vehicle decision factor* | | | |
| Involvement in decision on future household vehicle(s) | Sole decision-maker | 1,650 | 46.17 |
|  | Primary decision-maker | 868 | 24.29 |
|  | Equally share with other household member(s) | 1,056 | 29.55 |

## 5. Results

This section evaluates and interprets the estimation results of the SEM with latent variables (section 5.1) and the recursive trivariate model (section 5.2).

## 5.1. The estimated structural equation model with latent variables

*5.1.1. Model evaluation*

The estimated SEM with latent variables fits the data well based on the measures presented below Table 2. Specifically, the goodness-of-fit index (GFI) and the adjusted GFI are respectively equal to 0.965 and 0.944, which are greater than the critical value of 0.9 (Gao et al., 2017). Also, the standardized root mean square residuals (SRMR) and the root mean square of error approximation (RMEA) are respectively 0.049 and 0.047, which do not exceed the critical value of 0.05, according to Byrne (2016) for SRMR, and Steiger (1990) and Browne and Cudeck (1992) for RMEA. Moreover, the model chi-square is statistically significant at p-value < 0.001 (Golob, 2003).

*5.1.2. Model interpretation*

The estimated latent constructs reflect the individuals' preferences for four vehicle attributes (i.e., vehicle cost, reliability, performance, and refueling) and for shared mobility technologies. Table 2 lists the estimated weights (coefficients) of the latent constructs on their underlying observed indicators (denoted by $\boldsymbol{\omega}_r$ in the measurement equations (1) and discussed below) and the coefficients of exogenous variables explaining the latent constructs (denoted by $\boldsymbol{\tau}_l$ in the structural equations (2) and discussed below). During the estimation process, the coefficients were normalized between -1 and 1 to avoid identification problem. The reported t-statistic values indicate the statistical significance of almost all coefficients at the 5% level. As reported below Table 2, the latent propensity for shared mobility explains almost half of the model variance, while the remaining variance is explained in a descending order by the latent preferences for vehicle reliability, performance, refueling, and cost.

5.1.2.1. Measurement equations

Among the four latent constructs explaining the individuals' preferences for vehicle attributes, Table 2 shows that the latent variable pertaining to **vehicle cost** has positive loadings on its observed indicators (i.e., vehicle price and fuel economy). This is intuitive as cost-sensitive persons likely prefer lower vehicle cost, particularly the fixed cost component (vehicle price) which takes a greater load than the variable cost-related component (fuel economy). Second, the latent construct explaining **vehicle reliability** has positive impacts in descending magnitudes on the indicators inquiring about reliability, brand, and warranty, while it negatively contributes to the indicator describing seating capacity. This clearly implies that pro-reliability individuals are likely more attentive to the overall functional reliability of a vehicle than to the car manufacturers' reputation and marketing (through brand and warranty) and the vehicle physical size. Third, the latent propensity for **vehicle performance** places positive loads with diminishing scales on vehicle acceleration, horsepower, and towing capacity, indicating the priority of vehicle acceleration for pro-performance individuals. Finally, the latent preference for **vehicle refueling** has positive weights on all its indicators, with a greater contribution to the driving range of a fully fueled vehicle than to the time needed to refuel a vehicle and the fuel availability.

Furthermore, the latent propensity for **shared mobility** explains the individuals' preferences for existing mobility-on-demand services, which is inquired in terms of both the level of participation in carsharing and ridesharing programs as well as the expected impact of these technologies on the future decision about acquiring a vehicle. The associated positive weights in Table 2 suggest that pro-shared mobility individuals probably prefer these technologies and anticipate the associated influence on their next vehicle ownership decision.

5.1.2.2. Structural equations

The individuals' latent preferences for vehicle attributes and shared mobility are further tied to their socio-economic characteristics (Table 2). At the individual level, **gender** appears to be influential on the latent propensities for the four vehicle attributes. Everything else equal, women are likely more sensitive to vehicle cost, but less concerned about vehicle reliability, performance, and refueling, as compared to men. In addition, results indicate the role of **employment type** in shaping the latent tendencies toward both vehicle attributes and shared mobility. The cohort of full-time employees are potentially more disposed to be pro-reliability, pro-performance, and pro-shared mobility compared to others. While being more attentive to vehicle refueling, self-employed persons are also likely pro-shared mobility but to a lesser degree than full-time employees.

At the household level, persons living in **households with more children** (age < 12) have greater tendency towards vehicle performance and shared mobility but are less sensitive to vehicle cost compared to others. The preference for vehicle reliability is negatively associated with the number of adults and especially children in a household. This may not be surprising considering the underlying indicators of the latent vehicle reliability construct, among which the largest weight belongs to the vehicle seating capacity that likely receives greater attention by larger households. Finally, the logarithmic form of the number of adults in a household suggests a diminishing corresponding impact on the preference for vehicle reliability.

Table 2. Estimation results of the structural equation model with latent variables

|  | Latent preferences for … | | | | | | | | | |
|---|---|---|---|---|---|---|---|---|---|---|
|  | Vehicle cost | | Vehicle reliability | | Vehicle performance | | Vehicle refueling | | Shared mobility | |
|  | coef. | t-stat | coef. | t-stat | coef. | t-stat | coef. | t-stat | coef. | t-stat |
| **Measurement equations** | | | | | | | | | | |
| *Indicators of vehicle cost* | | | | | | | | | | |
| Vehicle price | 0.323 | 4.12 | | | | | | | | |
| Vehicle fuel economy (MPG) | 0.199 | 4.02 | | | | | | | | |
| *Indicators of vehicle reliability* | | | | | | | | | | |
| Reliability | | | 0.226 | 9.04 | | | | | | |
| Brand (vehicle make) | | | 0.094 | 4.24 | | | | | | |
| Warranty | | | 0.089 | 5.26 | | | | | | |
| Seating capacity | | | -0.662 | -11.95 | | | | | | |
| *Indicators of vehicle performance* | | | | | | | | | | |
| Acceleration | | | | | 0.397 | 5.01 | | | | |
| Horsepower | | | | | 0.178 | 3.98 | | | | |
| Towing capacity | | | | | 0.110 | 2.95 | | | | |

|  | Latent preferences for … | | | | | | | | | |
|---|---|---|---|---|---|---|---|---|---|---|
|  | Vehicle cost | | Vehicle reliability | | Vehicle performance | | Vehicle refueling | | Shared mobility | |
|  | coef. | t-stat | coef. | t-stat | coef. | t-stat | coef. | t-stat | coef. | t-stat |
| **Indicators of vehicle refueling** | | | | | | | | | | |
| Driving range | | | | | | | 0.334 | 4.66 | | |
| Refueling time | | | | | | | 0.136 | 3.31 | | |
| Fuel availability | | | | | | | 0.128 | 3.32 | | |
| **Indicators of shared mobility** | | | | | | | | | | |
| Level of using carsharing | | | | | | | | | 0.745 | 42.33 |
| Level of using ridesharing | | | | | | | | | 0.567 | 35.40 |
| Car/ridesharing affects future vehicle decisions | | | | | | | | | 0.604 | 34.56 |
| **Structural equations** | | | | | | | | | | |
| ***Socio-economic characteristics*** | | | | | | | | | | |
| *Gender* | | | | | | | | | | |
| Female | 0.205 | 3.76 | -0.057 | -2.40 | -0.210 | -4.30 | -0.336 | -4.46 | – | – |
| *Employment type* | | | | | | | | | | |
| Full-time | – | – | 0.045 | 1.94 | 0.160 | 3.51 | – | – | 0.270 | 14.47 |
| Self-employed | – | – | – | – | – | – | 0.099 | 1.92 | 0.074 | 3.70 |
| *Household structure* | | | | | | | | | | |
| Log (No. of adults) | – | – | -0.085 | -3.77 | – | – | – | – | – | – |
| No. of children | -0.106 | -2.17 | -0.352 | -9.99 | 0.141 | 2.94 | – | – | 0.164 | 7.22 |

*Notes:* Goodness-of-fit index (GFI) = 0.965, adjusted GFI = 0.944, standardized root mean square residuals (SRMR) = 0.049, and root mean square of error approximation (RMEA) = 0.047. The explained variance by the five latent constructs are 6.44% (vehicle cost), 22.58% (vehicle reliability), 9.00% (vehicle performance), 6.54% (vehicle refueling), and 55.44% (shared mobility). "–" denotes a statistically insignificant parameter that is removed from the model.

## 5.2. The estimated recursive trivariate model

### 5.2.1. Model evaluation

The proposed recursive trivariate model is estimated using the quasi-newton optimization method with 70 random draws and converged after 187 iterations. Table 3 presents the results. As stated earlier in section 1 and mathematically expressed in Eqs. (3)-(5), the main hypothesis of this study pertains to the presence of endogeneity originating from the structural effects of: 1) the concern about AV safety on the AV acceptance, and 2) the current annual VMT on both the AV acceptance and concern about AV safety. More clearly, we hypothesize that these three endogenous outcomes should be modeled jointly (i.e., the contemporaneous or cross-equation error correlations $\rho_{12}$, $\rho_{13}$, and $\rho_{23}$ in Eq. (6) should not be constrained to be zero) in order to correctly explain each endogenous outcome using the other endogenous outcomes. Again, this is because the joint estimation allows for absorbing any propensity for the endogenous outcomes that is associated with omitted or unobserved factors (e.g., lifestyle, personal values, etc.) that have common effects on the outcome variables.

To test the above hypothesis, we firstly note in Table 3 that the estimated unobserved error correlations are highly statistically significant clearly confirming the hypothesis. We further evaluate the hypothesis by comparing the estimated recursive trivariate model with a similarly specified but restricted model corresponding to an independent sequence of three disjoint (univariate) equations, wherein all error correlations are constrained to be zero (i.e., $\rho_{12} = \rho_{13} = \rho_{23} = 0$). Comparing the log-likelihood values at convergence of -16,905 (for recursive trivariate model) and -17,078 (for univariate model counterparts), the likelihood ratio test (Greene, 2000; Greene and Hensher, 2010; Washington et al., 2003) yields a value of 346 (= -2(-16905-17078)). This is greater than the associated chi-square table value at p-value << 0.001, again corroborating the hypothesis. Consequently, ignoring the correlated unobserved factors could lead to erroneous estimates of the model parameters and potentially misleading policy implications. Specifically, the positive signs of the three estimated error correlations in Table 3 imply that the common unobserved factors influence the endogenous outcomes in the same direction. Hence, the structural associations of the endogenous outcomes (denoted by $\theta_{12}$, $\theta_{13}$, and $\theta_{23}$ in Eqs. (4)-(5)) might be artificially underestimated if the effect of shared unobserved factors is ignored. This is evidenced by comparing the corresponding estimated coefficients in the recursive trivariate model ($\theta_{12} = -0.171$, $\theta_{13} = -0.233$, $\theta_{23} = -0.661$) and the univariate counterparts ($\theta_{12} = -0.170$, $\theta_{13} = -0.230$, $\theta_{23} = -0.392$). The behavioral interpretation of these structural effects is discussed in section 5.2.2.

Note that, as implied above, the common unobserved error correlations (which relate to endogeneity) should be distinguished from the pure structural interdependencies among the endogenous outcomes (which relate to recursivity). With the presence of the former tested and confirmed above, it is also worth testing the latter. Table 3 shows that the coefficients capturing the linear associations among the three endogenous outcomes (i.e., $\theta_{12}$, $\theta_{13}$, and $\theta_{23}$ in Eqs. (4)-(5)) are highly statistically significant, thus confirming the recursive structural system. The recursivity is also assessed using the likelihood ratio test by comparing the recursive trivariate model with a restricted trivariate model absent recursivity (i.e., $\theta_{12} = \theta_{13} = \theta_{23} = 0$). The test gives a chi-square value of 322 (= -2(16905-17066)) at p-value << 0.001, which clearly rejects the restricted model in this particular empirical context.

Furthermore, the estimated recursive trivariate model outperforms its univariate model counterparts in terms of model fit and explanatory power. The better fit of the proposed model can be verified by the larger value of $\rho_c^2 = 0.025$ compared to $\rho_c^2 = 0.013$ for the univariate counterparts. Also, the proposed model has smaller values of the Akaike information criterion (AIC) and the Bayesian information criterion (BIC), which are respectively equal to 9.483 and 9.554, than those of the univariate model counterparts (AIC = 9.602 and BIC = 9.662), thereby indicating the better fit of the proposed model (Greene, 2000). Moreover, the recursive trivariate model has a superior explanatory power compared to the univariate model counterparts as some of the statistically significant explanatory variables in the proposed model are found to be statistically insignificant in the univariate models (e.g., the variables explaining vehicle reliability, white individuals, self-employed individuals, and households with low annual income).

### 5.2.2. Model interpretation

To streamline the model interpretation (Table 3), below we discuss in sequence the structural dependence among the endogenous outcomes ($\theta$'s in Eqs. (4)-(5)), and the impacts on the three endogenous outcomes of the latent preferential variables ($\boldsymbol{\beta}$'s in Eqs. (3)-(5)) and exogenous variables ($\boldsymbol{\alpha}$'s in Eqs. (3)-(5)). Note that, as discussed earlier in section 3, the estimation results of an ordinal-outcome model can be only partially interpreted. Specifically, a positively signed coefficient indicates that an increase in the associated explanatory variable increases the probability of the highest ordinal level (i.e., strongly agree) and decreases the probability of the lowest ordinal level (i.e., strongly disagree). The converse holds true for the negative coefficients. To interpret the three intermediate ordinal levels (i.e., agree, neutral, and disagree), the marginal effect of each level is calculated using Eqs. (18)-(19) and presented in Figure 5. The marginal effect values refer to the approximate changes in the probability of each agreement level in response to a unit change in the desired explanatory variable while other variables are held constant at their respective sample mean.

5.2.2.1. Interdependencies among endogenous outcomes

As illustrated in section 5.2.1, the structural endogeneity effects of the three outcome variables are accommodated through capturing the correlated unobserved factors $\rho_{12}$, $\rho_{13}$, and $\rho_{23}$ that jointly influence the endogenous outcomes. With the structural endogeneity controlled, we can now interpret the true structural relationships among the endogenous outcomes, which are purified from the outcome correlations caused by the common unobserved factors. These structural relationships are represented by the coefficients $\theta_{12}$, $\theta_{13}$, and $\theta_{23}$ in Eqs. (4)-(5) and the solid red and blue arrows in Figure 1.

Foremost, the negative sign of $\theta_{23}$ (-0.661) in Table 3 indicates the negative structural effect of the individuals' **concern about AV safety** on their inclination toward **AV acceptance**. More exactly, those who have greater concern about AV safety will more likely "strongly" disfavor the AV technology. This finding is also corroborated by the prior work on, for instance, choosing AV as last mile mode to public transit (Yap et al., 2016), owning versus using on-demand AVs (Nazari et al., 2018), perception of autonomous public transit (Kassens-Noor et al., 2020), and AV acceptance (Xu et al., 2018; Zhang et al., 2019). Also looking at the marginal effect values in Figure 5, it is further revealed that one unit growth in an individual's concern about AV safety will boost the probability of strong disagreement with AV acceptance by 0.196, which is more tangible than the corresponding drop in the probability of strong agreement with AV acceptance (by 0.140). These asymmetrical marginal effect values of AV safety concern on the two opposing strong levels of (dis)agreement with AV acceptance, however, conversely hold for the related moderate (dis)agreement levels. That is, a unit increase in safety concern will likely have a greater influence on the moderate agreement (which is reduced by 0.048) than the moderate disagreement (which is trivially increased by 0.010) with AV acceptance. In addition, the relative magnitudes of these marginal effects suggest that policies aiming to promote AV acceptance through diminishing AV safety concern could be much less effective for the cohort of individuals who are neutral about or moderately against AV acceptance, as compared to those who strongly (dis)like or moderately welcome AVs. More importantly, AV safety concern has the second-largest marginal effects (in absolute values) on all (dis)agreement levels of AV acceptance compared to other explanatory variables, only next to those of the latent preferences for vehicle cost (discussed in section 5.2.2.2). This unveils the room for advancing the public intention to accept the AV technology through, for instance, educational policy interventions advancing the public familiarity with and trust in AV safety.

As for the current annual **VMT**, we find its negative structural effect on the public tendency towards **AV acceptance**, as signified by the negative value of the respective coefficient $\theta_{13}$ (-0.233). Interestingly, this finding reveals that persons who drive more (i.e., greater VMT) will more likely "strongly" disapprove AVs, which is also found by prior works such as the study of Haboucha et al. (2017) on preferences for AV as a travel mode. Since larger-VMT individuals probably spend more time in vehicle, their strong disinclination to accept AVs might challenge the assumption usually made in prior studies about the AV riders' perceived benefit from in-vehicle time saving due to engagement in non-driving activities (Steck et al., 2018; Rashidi et al., 2020; Szimba and Hartmann, 2020), and thus may also question the consequent potential of AVs in urban sprawl (Meyer et al., 2017; Haboucha et al., 2017). Furthermore, the relevant marginal effects in Figure 5 indicate that a unit rise in the logarithm of the current annual VMT will change the probabilities of strong agreement (by -0.049) and strong disagreement (by 0.069) with AV acceptance, whereas the associated marginal effects on the moderate and neutral levels of AV acceptance are trivial. Note that the logarithmic appearance of the current annual VMT highlights that the above-mentioned impacts are diminishing for larger VMT values. The above discussion suggests that policies aiming to popularize AVs should center their efforts on lower-VMT individuals including, among others, persons with travel-restrictive conditions.

The negative association of current annual **VMT** with AV acceptance becomes even more compelling by noting that the AV acceptance of the larger-VMT cohort is likely hindered more by factors other than the **concern about AV safety**, as the latter was found (through the negative sign of $\theta_{12} = -0.171$) to likely less "strongly" worry such individuals. This is worth noting, since one may conjecture that larger-VMT travelers should be more worried as they spend more time in vehicle and thus are probably more frequently exposed to possible AV malfunctions or cyber-security issues affecting their safety. In addition, the marginal effects of VMT on AV safety concern (Figure 5) are found to be relatively small, except for the associated effect on the strong level of AV safety concern whose probability will reduce by 0.064 if the logarithm of VMT increases by one unit. Thus, policies advocating safe autonomous mobility could focus on lower-VMT individuals, though they will be more effective in alleviating the "strong" level of safety concern.

5.2.2.2. Effects of latent constructs on endogenous outcomes

Three of the latent preferential variables estimated in section 5.1 are found to statistically significantly describe the intent to accept AVs and the concern about AV safety (Table 3). First and foremost, individuals who are more sensitive to **vehicle cost** will likely strongly not intend to accept AV, which is not surprising as the high-tech AVs might be appraised at a higher price than the conventional human-driven vehicle technology (Howard and Dai, 2014; Haboucha et al., 2017; Noruzoliaee et al., 2018; Wang et al., 2021). Therefore, vehicle cost-conscious persons are probably not among the early adopters of AVs, given the anticipated higher price of AV technology at least at its early stage of commercial availability. More notably, comparing the marginal effects of all explanatory variables in Figure 5, it is revealed that a unit change in the latent vehicle cost sensitivity would cause the greatest changes in all five levels of agreement with AV acceptance. In particular, a unit increase (decrease) in vehicle cost sensitivity would decrease (increase) the probability of strong agreement (strong disagreement) with AV acceptance by 0.215 (0.301).

In addition, the latent preferences for **vehicle reliability** and **shared mobility** can mitigate the public concerns about the safety of AV technology. More exactly, the pro-reliability and pro-shared mobility individuals do not retain a firm concern about AV safety, as implied by their negative coefficients in Table 3. This may be explained by the potentially better familiarity of pro-reliability individuals with technical aspects of vehicles in general (and thus their understanding of how safely AVs can indeed operate), as well as by the mobility tech-savviness of pro-shared mobility individuals. It is also worth noting that a unit change in the preferences for vehicle reliability and shared mobility will induce the greatest changes in the probabilities of all five levels of AV safety concern compared to those caused by all other explanatory variables. A unit increase in the latent preferences for vehicle reliability and shared mobility will shrink the probability of being strongly safety-concerned by 0.094 and 0.097, respectively.

5.2.2.3. Effects of exogenous explanatory variables on endogenous outcomes

The socio-economic, residential, and vehicle decision attributes are also found to shape the individuals' intention to accept AVs, safety concern about the AV technology, and current annual VMT. Among the individual-level socio-economic characteristics, gender, race, education level, and employment type explain the three outcome variables. Specifically, **women** are likely strongly concerned about the safety of driverless cars compared to men, as evidenced by the associated positive coefficient in Table 3. A similar behavior is found by Dannemiller et al. (2021) who label it as "gender gap in AV safety concern" which is also previously verified in science, technology, engineering, and mathematics (STEM) (Wang and Degol, 2017). Among various **race** categories specified earlier in Table 1, White individuals are potentially less safety-concerned and drive more (i.e., larger VMT), whereas Asians will more likely intend to accept AVs. Focusing on the **education level**, those with postgraduate degree or higher are probably less concerned about AV safety, which may be attributable to their potentially more extensive acquaintance with emerging technologies. Moreover, persons with any level of postsecondary education drive more (in terms of VMT) than lower-educated individuals do. However, VMT does not monotonically increase with the level of postsecondary education, as obvious from the associated larger coefficient of college graduates (those with 4-year degree) than postgraduates. As for **employment type**, the estimation results reveal the positive tendency of full-time employees and self-employed persons for AV acceptance. Similarly, Wang et al. (2021) report the positive tendency of full-time workers towards AV adoption. Besides, full-time employees generate greater VMT than others do, possibly owing to their commute trips.

Turning to the household-level socio-economic factors, **household structure** appears in the model as two continuous variables indicating the numbers of children (age<12) and teenage children (12≤age<16) in a household. Those who live in households with more children or teenage children, ceteris paribus, show a greater penchant for AV acceptance, with the associated effect of the presence of teenage children being more pronounced. This is not surprising, since the presence of more (teenage) children could translate into the generation of more trips, which can be more effectively served by driverless cars. **Household income** also affects the proclivity for AV acceptance and VMT in intuitive ways. Households earning less than $75k annually more likely do not embrace AVs and also travel less than others, whereas those with annual income exceeding $200k are more inclined to accept AVs as also reported in Bansal et al. (2016).

Table 3. Estimation results of the recursive trivariate model

| | AV acceptance | | AV safety concern | | Log (VMT) | |
|---|---|---|---|---|---|---|
| | coef. | t-stat | coef. | t-stat | coef. | t-stat |
| Constant | 5.378 | 10.95 | 3.270 | 7.05 | 7.681 | 46.32 |
| **Endogenous variables** | | | | | | |
|   AV safety concern | -0.661 | -12.13 | | | | |
|   Log (VMT) | -0.233 | -2.80 | -0.171 | -2.83 | | |
| **Latent variables** | | | | | | |
|   Vehicle cost | -1.019 | -2.82 | – | – | – | – |
|   Vehicle reliability | – | – | -0.251 | -2.18 | – | – |
|   Shared mobility | – | – | -0.259 | -9.87 | – | – |
| **Exogenous variables** | | | | | | |
| *Socio-economic characteristics* | | | | | | |
| *Gender* | | | | | | |
|   Female | – | – | 0.198 | 5.71 | – | – |
| *Race* | | | | | | |
|   White | – | – | -0.069 | -1.94 | 0.124 | 2.08 |
|   Asian | 0.239 | 4.68 | – | – | – | – |
| *Education level* | | | | | | |
|   Below college graduate | – | – | – | – | 0.291 | 2.49 |
|   College graduate | – | – | – | – | 0.445 | 3.48 |
|   Postgraduate or higher | – | – | -0.096 | -2.66 | 0.374 | 3.04 |
| *Employment type* | | | | | | |
|   Full-time employed | 0.096 | 2.38 | – | – | 0.270 | 4.28 |
|   Self-employed | 0.115 | 1.90 | – | – | – | – |
| *Household structure* | | | | | | |
|   No. of children (age < 12) | 0.067 | 2.92 | – | – | – | – |
|   No. of teenage children (12 ≤ age < 16) | 0.132 | 2.97 | – | – | – | – |
| *Household income* | | | | | | |
|   Low (< $75k per year) | -0.098 | -2.60 | – | – | -0.104 | -1.96 |
|   High (≥ $200k per year) | 0.155 | 3.55 | – | – | – | – |
| *Residential factors* | | | | | | |
| *Parking cost at residence* | | | | | | |
|   Free parking | -0.153 | -1.91 | – | – | 0.473 | 4.04 |
|   Daily parking cost ($/day) | – | – | – | – | -0.021 | -5.41 |
| *Vehicle decision factor* | | | | | | |
| *Involvement in decision on future household vehicle(s)* | | | | | | |
|   Sole decision-maker | – | – | – | – | -0.188 | -3.24 |
|   Equally shared with others | -0.150 | -3.49 | -0.112 | -2.70 | – | – |
| *Error Correlations* | | | | | | |
|   AV acceptance | | | | | | |
|   AV safety concern | 0.530 | 8.38 | | | | |
|   Log (VMT) | 0.352 | 2.40 | 0.248 | 2.32 | | |
| *Thresholds (of ordinal variables)* | | | | | | |
|   Threshold 2 | 0.432 | 13.16 | 0.488 | 14.01 | | |
|   Threshold 3 | 0.931 | 14.32 | 1.018 | 21.49 | | |
|   Threshold 4 | 1.552 | 14.74 | 1.878 | 28.61 | | |

*Notes:* Threshold 1 is fixed at zero. $LL(\beta) = -16{,}905$, $\rho_c^2 = 0.025$, $AIC = 9.483$, and $BIC = 9.554$. "–" denotes a statistically insignificant parameter that is removed from the model.

The residential neighborhood of an individual, represented by the **parking cost at residence**, influences his/her intention to accept AV and VMT. Interestingly, those with free parking at residence are more likely dissuaded to accept AVs as also implied by Liljamo et al. (2018), possibly due to the less tangible benefit accrued from the self-parking capability of AV in the absence of residential parking cost constraint. Moreover, free residential parking is associated with more driven miles, likely because free parking (such as private garage or driveway) is more commonplace in less dense neighborhoods (e.g., suburbs) that are possibly distant from activity locations. In contrast, those incurring higher residential parking cost drive less.

Finally, stronger intra-household interactions in making the household's future **vehicle decision** are likely negatively associated with AV acceptance and concern about AV safety. More clearly, compared to individuals who take the role of a sole or primary player in the future vehicle decision of their household, those who share the decision with other household members are likely less worried about the AV safety, less inclined towards AV acceptance, and driver more (i.e., larger VMT) than others.

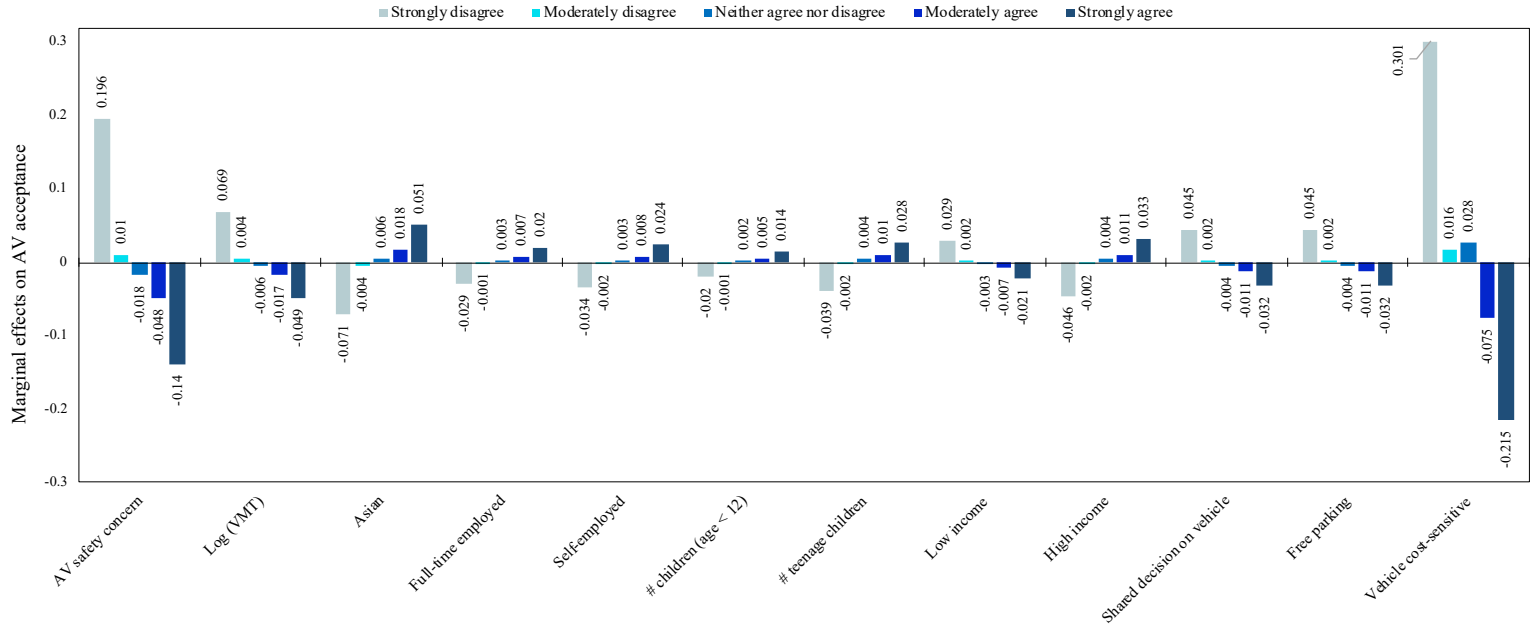

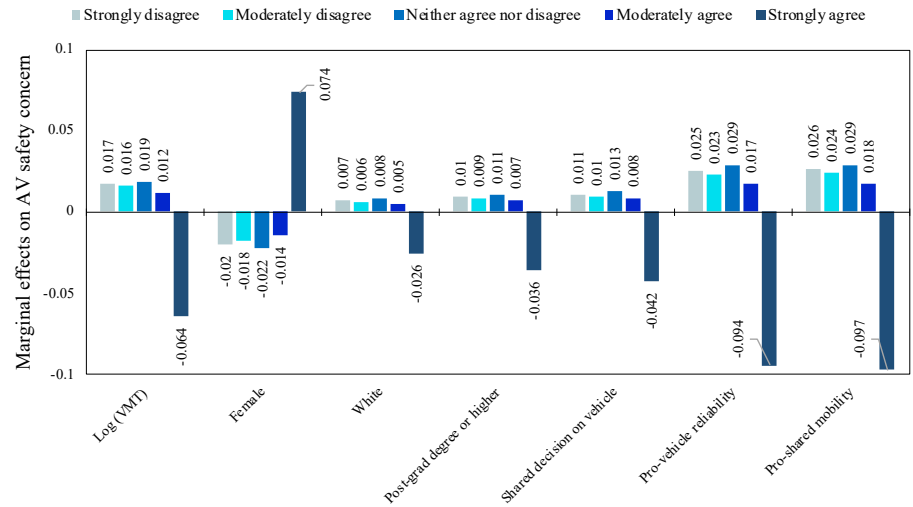

**Figure 5. Marginal effects for the recursive trivariate model with ordinal-continuous outcomes**

# 6. Conclusions and policy implications

Despite the maturing road tests and restricted commercial mobility services with automated vehicles (AVs), the existing behavioral research and surveys suggest that, to date, the public is largely reluctant or neutral to accept this emerging technology. The persistence of this demand landscape for AVs, however, could curb the promising economic, societal, and environmental benefits of prevalent autonomous mobility. Proactive policy interventions are thus needed early on to provide impetus for AV acceptance, which should be informed by an in-depth understanding of the *AV acceptance behavior* of the public in order to identify the determinants thereof and direct the policies towards appropriate population groups. Among the potential determinants, this paper particularly attempts to unravel the roles of *perceived concern about AV safety* and *current (pre-AV era) travel behavior*, the latter approximated by the current annual vehicle-miles traveled (VMT). The joint estimation of the three outcomes in a recursive trivariate econometric model allows to uncover the "true" interdependencies among them, cleansed of any correlated unobserved factors that can have common impacts on these endogenous outcomes. Moreover, since AV acceptance behavior is related to vehicle and mobility decisions, the public latent preferences for four vehicle attributes (namely, vehicle cost, reliability, performance, and refueling) and for existing shared mobility systems are also accounted for using a structural equation model (SEM).

The SEM and the recursive trivariate model are estimated in sequence on a stated preference survey in the State of California to explore how Californians would likely react to the advent of AVs. The estimation results indicate that the AV acceptance behavior is influenced most — through having the largest marginal effects on AV acceptance compared to those of other explanatory factors — by the sensitivity for vehicle cost and the perceived concern about AV safety, both hindering AV acceptance. This unveils the room for stimulating AV acceptance through policy levers such as cost incentive programs (e.g., lower interest financing for AVs compared to conventional cars, cash rebates, and dealer marketing supports) and trust-building programs (e.g., technology demonstration and public educational campaigns in the media aiming to acquaint with and build trust in AV safety), respectively. These policies can be more effectively designed by targeting towards: 1) facilitating the utilization of existing shared mobility systems, which is found to significantly alleviate the perceived concern about AV safety (through having the largest marginal effects on AV safety concern), and might be justified by the potential tech-savviness of pro-shared mobility users; 2) advancing vehicle reliability awareness, which has the second-largest marginal effects on mitigating the AV safety concern, and could be explained by the better familiarity of pro-reliability individuals with technical aspects of vehicles in general and thus their understanding of how safely AVs could indeed operate; and 3) closing the gender gap, since females are found to be likely considerably cost-conscious and safety-concerned (through having the third-largest marginal effects on AV safety concern).

To a lesser degree than the vehicle cost sensitivity and AV safety concern discussed above, the policy interventions aiming to diffuse AV acceptance should also be centered on bridging the racial gap and appreciating the diversity of current travel behavior (i.e., current annual VMT). Compared to other races, Asians are found to be more enthusiastic about the AV technology. Interestingly, those who currently drive more (i.e., larger annual VMT) will likely be less disposed to AV acceptance, despite being less concerned about AV safety. Because larger-VMT individuals probably spend more time in vehicle, their lower inclination towards AV acceptance might challenge the conjectures about AV riders' benefit from in-vehicle time saving due to engagement in non-driving activities, and thus may also question the consequent potential of AVs in urban sprawl. Finally, it is worth noting that all the above-discussed policy insights geared towards popularizing AVs are most effective (through showing the largest marginal effects) on reshaping the opinions of individuals with "strong" (dis)agreement with AV acceptance. By contrast, those with "neutral" and "moderate" (dis)agreement with AV acceptance, are found to be more conservative and retain firm opinions, thus rendering the above policies less effective for them.

Future research can be extended in two main directions to complement the present study. From the modeling perspective, more advanced yet exceedingly computationally expensive econometric models can be employed to consider unobserved heterogeneity through random parameters, non-fixed ordinal thresholds, and integration of the SEM into the recursive trivariate model. In the second direction, the future studies are suggested to collect more recent and richer databases to yield a more realistic model of the public's opinion about AVs. On updating the time of data collection, since the public is constantly exposed to new information about the AV technology, collecting a more recent dataset can advance our understanding of the public's latest opinion about AV acceptance. On enhancing the data collection method, experimental design tools can be employed to collect respondents' opinion about AVs in a controlled stated preferences setting by providing detailed description of AV attributes, such as purchase cost and travel time, in comparison with conventional human-driven vehicles.

## Acknowledgment

This research was funded in part by the U.S. Department of Transportation through the Safety21 National University Transportation Center (No. 69A3552344811/69A3552348316). This research was also funded in part by the US National Science Foundation (Award No. 2112650). The opinions expressed are solely those of the authors, and do not necessarily represent those of the USDOT and NSF. The authors are greatly thankful to National Renewable Energy Laboratory and California Energy Commission for providing the database.

18. Fagnant, D.J., Kockelman, K.M., 2014. The travel and environmental implications of shared autonomous vehicles, using agent-based model scenarios. Transportation Research Part C: Emerging Technologies 40, 1–13. https://doi.org/10.1016/j.trc.2013.12.001

19. Favarò, F.M., Nader, N., Eurich, S.O., Tripp, M., Varadaraju, N., 2017. Examining accident reports involving autonomous vehicles in California. PLoS one 12, e0184952. https://doi.org/10.1371/journal.pone.0184952

20. Fraedrich, E., Heinrichs, D., Bahamonde-Birke, F.J., Cyganski, R., 2019. Autonomous driving, the built environment and policy implications. Transportation Research Part A: Policy and Practice 122, 162–172. https://doi.org/10.1016/j.tra.2018.02.018

21. Gao, Y., Rasouli, S., Timmermans, H., Wang, Y., 2017. Understanding the relationship between travel satisfaction and subjective well-being considering the role of personality traits: A structural equation model. Transportation Research Part F: Traffic Psychology and Behaviour 49, 110–123. https://doi.org/10.1016/j.trf.2017.06.005

22. Gkartzonikas, C., Gkritza, K., 2019. What have we learned? A review of stated preference and choice studies on autonomous vehicles. Transportation Research Part C: Emerging Technologies 98, 323–337. https://doi.org/10.1016/j.trc.2018.12.003

23. Golob, T.F., 2003. Structural equation modeling for travel behavior research. Transportation Research Part B: Methodological 37, 1–25. https://doi.org/10.1016/S0191-2615(01)00046-7

24. Gomes, L., 2014. Hidden obstacles for Google's self-driving cars: Impressive progress hides major limitations of Google's quest for automated driving. Retrieved from: https://www.technologyreview.com/2014/08/28/171520/hidden-obstacles-for-googles-self-driving-cars/. Massachusetts Institute of Technology.

25. Greene, W.H., 2000. Econometric analysis. Prentice-Hall, New York, NY, USA.

26. Greene, W.H., Hensher, D.A., 2010. Modeling ordered choices: A primer. Cambridge University Press.

27. Haboucha, C.J., Ishaq, R., Shiftan, Y., 2017. User preferences regarding autonomous vehicles. Transportation Research Part C: Emerging Technologies 78, 37–49. https://doi.org/10.1016/j.trc.2017.01.010

28. Harb, M., Stathopoulos, A., Shiftan, Y., Walker, J.L., 2021. What do we (Not) know about our future with automated vehicles? Transportation Research Part C: Emerging Technologies 123, 102948. https://doi.org/10.1016/j.trc.2020.102948

29. Harb, M., Xiao, Y., Circella, G., Mokhtarian, P.L., Walker, J.L., 2018. Projecting travelers into a world of self-driving vehicles: Estimating travel behavior implications via a naturalistic experiment. Transportation 45, 1671–1685. https://doi.org/10.1007/s11116-018-9937-9

30. Harper, C.D., Hendrickson, C.T., Mangones, S., Samaras, C., 2016. Estimating potential increases in travel with autonomous vehicles for the non-driving, elderly and people with travel-restrictive medical conditions. Transportation Research Part C: Emerging Technologies 72, 1–9. https://doi.org/10.1016/j.trc.2016.09.003
31